\documentclass[11pt,a4paper]{article}
\usepackage[utf8]{inputenc}
\usepackage[T1]{fontenc}
\usepackage{amsmath,amssymb}
\usepackage{booktabs}
\usepackage{graphicx}
\usepackage{geometry}
\usepackage{natbib}
\usepackage{hyperref}
\usepackage{siunitx}
\usepackage{caption}
\usepackage{placeins}
\hypersetup{colorlinks=true,linkcolor=blue,citecolor=blue,urlcolor=blue,
pdftitle={Earnings Price Discovery Outside Regular Hours: Evidence from S\&P 1500 Releases, 2016-2026},pdfauthor={Philipp D. Dubach},
pdfkeywords={earnings announcements, extended-hours trading, price discovery, weighted price contribution, market microstructure}}
\graphicspath{{figures/}}

\title{Earnings Price Discovery Outside Regular Hours:\\[2pt]
\large Evidence from S\&P 1500 Releases, 2016--2026}

\author{Philipp D.~Dubach\\[2pt]
\normalsize\textit{Zurich, Switzerland}\\
\normalsize\href{mailto:phdubach@pm.me}{phdubach@pm.me}}
\date{29 November 2026}

\begin{document}
\maketitle

\begin{abstract}
\noindent
Trading outside regular hours has grown over the past decade. We test whether it now accounts for
a larger share of the price response to earnings news. The sample contains 29{,}907 earnings
releases by S\&P 1500 firms from December 2016 to September 2026, measured with 1-minute bars from
a free data source and with official opening prices. For releases after the close, the
after-hours session accounts for 0.42 of the event return. Its trend is $+0.0007$ per year
(standard error 0.0052). The implied change over the decade, $+0.04$, is not statistically
significant; the upper bound of its \SI{95}{\percent} confidence interval is 0.10, or 0.13 under
alternative benchmarks. In the data, the after-hours share of event volume remains between 7 and
\SI{10}{\percent}, while its share of trades nearly doubles. The intraday bars alone indicate that
the next-morning share of the response doubled, and this increase passes every check based on the
bars. Half of it, however, reflects stale pre-market prices in the early years: measured to the
official open, the increase is 0.02 and not significant. For releases after the close, the share
of the response priced by the official open fell in 2020--2023 and then recovered. In 2024--2026
it exceeds its 2016--2019 level by 0.05 to 0.07, close to the increase in the same share on days
without earnings news. Small firms show the same trend as large firms, and the opening price
predicts the next close as well in 2026 as in 2017. None of the three hypotheses is supported as
stated. The tests can detect an increase of about 0.1 over the decade, but not a much smaller one.
\end{abstract}



\section{Introduction}

Trading outside regular hours has grown over the past decade. Retail brokers allow app users to
trade before the open and after the close, and overnight venues now trade listed stocks between
20:00 and 04:00 ET. A staff memo of the SEC's Division of Trading and Markets puts overnight volume
at \SI{0.9}{\percent} of NMS share volume in August 2026 \citep{sec-2026-overnight}. Most earnings
news arrives in these hours: in our sample period, \SI{92}{\percent} of the Item 2.02 filings by
index members come before the open or after 16:05 ET.

If more investors trade when the news is released, a larger part of the price response to
earnings should occur outside regular hours. We test three hypotheses:
\begin{itemize}
\item[H1:] the after-hours share of earnings price discovery increased from 2016 to 2026;
\item[H2:] the increase is larger for S\&P 500 firms than for S\&P 600 small caps, whose
after-hours trading remains thin;
\item[H3:] a larger part of the response is priced by the 09:30 open, so that less of it is left
to the regular session.
\end{itemize}
The hypotheses come from a one-page research plan dated 22 November 2026, written before any data
were downloaded and available from the author. The exact tests were specified on 28 November,
after the year-by-year shares were known, and the checks in Table~\ref{tab:robust2} were added
after the main results. The tests based on official opening prices (Table~\ref{tab:official}) were
added on 29 November during a second review, and they change the interpretation of the first
draft. The study is not pre-registered, and we state the origin of each test.

None of the hypotheses is supported as stated. For releases after the close, the after-hours
session accounted for 0.52 of the event return in 2017, 0.38 to 0.41 in 2020--2023 and 0.43 in
2026. With firm fixed effects and controls, its trend is $+0.0007$ per year with a standard error
of 0.0052. The implied change over the decade is $+0.037$, with a \SI{95}{\percent} confidence
interval from $-0.027$ to $+0.101$. We can reject neither zero nor an increase of a quarter of the
0.42 level. Under alternative benchmarks, the change from 2016--2019 to 2024--2026 is about $+0.04$
and significant at the 5 or \SI{10}{\percent} level; when the move to the closing auction is
excluded, it is $+0.01$. After-hours volume did not grow either. The after-hours share of event
volume ranges from 6.8 to \SI{9.6}{\percent} with no trend, while its share of trades rises from
4.2 to \SI{7.8}{\percent}. The after-hours session thus attracted more, but smaller, trades without
accounting for a much larger part of the price response.

The intraday bars alone suggest a second result, which does not survive a comparison with official
prices. Measured to the data's last trade before 09:30, the next-morning share of the response to
releases after the close rises from 0.09 in 2017 to 0.17 in 2026. In the early years, however, the
data contain few pre-market trades, so this last trade is often stale and the move of the morning
is assigned to the regular session. Official opening prices are not subject to this problem.
Measured to the official open, the next-morning share is 0.20 to 0.22 in 2018--2019 and in
2024--2026, and its change within firms is $+0.016$ (standard error 0.017).

For releases after the close, the share of the response priced by the official open was 0.65 in
2017--2019, fell to 0.57 in 2020--2023 and returned to 0.66 in 2024--2026. Within firms and with
controls, the share in 2024--2026 exceeds that of 2016--2019 by 0.049 (0.022) for releases after
the close and by 0.072 (0.019) for releases before the open. This increase is not specific to
earnings announcements. On days without earnings news, the share of a stock's daily return that is
priced by the open rose by about 0.04 over the same years.

Small firms follow the same trend as large firms. The opening price explains 52 to
\SI{72}{\percent} of the variance of the event return in every year, with no trend. Both tests
have limited power: the H2 test can detect a difference in trends of about 0.024 per year, and the
H3 test a change in $R^2$ of about 0.02 per year.

Earlier work measures the session shares of earnings price discovery for 2004--2008
\citep{jiang-2012-afterhours} and studies overnight trading in one quarter of 2024
\citep{eaton-2025-nocturnal}. This paper provides a year-by-year series for 2016--2026 across the
S\&P 1500 and separates the after-hours evening from the next morning. It also shows how changes
in the coverage of a free data source can resemble a change in the market: an increase in the
next-morning share that passes every check based on the bars falls by half when measured against
official opening prices. Scripts reproduce every number from the downloaded data.

Section~\ref{sec:literature} reviews the related literature. Section~\ref{sec:data} describes the
data and the sample, Section~\ref{sec:method} the methodology and Section~\ref{sec:results} the
results. Section~\ref{sec:discussion} discusses the limitations, Section~\ref{sec:conclusion}
concludes, and Appendix~\ref{app:validation} reports the validation of the data.

\section{Related literature}\label{sec:literature}

\paragraph{Price discovery outside regular hours.} \citet{barclay-warner-1993} introduce the
weighted price contribution to measure which trades move prices. \citet{barclay-hendershott-2003}
apply it to trading sessions and find that after-hours trades, although few, account for a large
part of price discovery, and \citet{barclay-hendershott-2004} show that thin after-hours trading
raises trading costs and the share of informed traders. \citet{cao-2000-preopening} and
\citet{barclay-hendershott-2008} study price discovery in the Nasdaq pre-open with and without
trading, and \citet{biais-1999-preopening} test whether pre-open prices predict the opening price,
which is the basis of our H3. \citet{hasbrouck-1995} measures each market's share of price
discovery with information shares, and \citet{vanbommel-2011} compares the weighted price
contribution with $R^2$ and variance-ratio measures.

\paragraph{Earnings news outside regular hours.} \citet{patell-wolfson-1982} show that firms tend
to release bad news after the close, and \citet{francis-1992} compare the price response to
earnings released during and outside trading hours. \citet{greene-watts-1996} compare how NYSE and
Nasdaq prices incorporate news released overnight and during the day, and
\citet{masulis-shivakumar-2002} find that the speed of the response to news released outside
regular hours depends on the listing venue. \citet{berkman-truong-2009} show that by the 2000s most
firms released earnings outside regular hours, so that the first trading day after the release is
the relevant event day. \citet{jiang-2012-afterhours} measure the share of the earnings response in
each session for 2004--2008. Firms choose the timing of their releases, which
\citet{doyle-magilke-2009}, \citet{dehaan-2015} and \citet{michaely-2016} study, and
\citet{dellavigna-pollet-2009} find slower responses to Friday releases.
\citet{christensen-2025-jumps} examine after-hours price jumps after earnings for 50 large stocks
up to 2020, and \citet{martineau-2022} shows that the prices of large firms now incorporate
earnings news within the announcement window.

\paragraph{Overnight returns and retail trading.} Overnight returns differ from intraday returns
\citep{berkman-2012,lou-polk-skouras-2019,hendershott-2020,boyarchenko-2023}. Retail investors
trade more on attention \citep{barber-2022-robinhood} and can be identified in trade data
\citep{boehmer-2021-retail}, and outages at retail brokers show that their trading affects market
quality \citep{eaton-2022-outages}. \citet{eaton-2025-nocturnal} study the new overnight venues in
one quarter of 2024.

\paragraph{Measurement.} \citet{gregoire-martineau-2022} study how after-hours trades and quotes
incorporate earnings news into prices; our data contain trades only. \citet{hu-stephan-2022} show
that releases at the closing bell can move the closing price, so we exclude releases from 16:00 to
16:05. \citet{bogousslavsky-muravyev-2023} document the growth of the closing auction, and one of
our checks separates its move from the after-hours window. Trade sizes have declined across the
market, and odd lots account for a growing part of trades \citep{ohara-2014-oddlots}.
\citet{lyle-stephan-yohn} find that initial reactions differ between releases before the open and
after the close, so we analyse the two timings separately throughout. \citet{rogers-2017-edgar}
show that the time at which a filing reaches investors can differ from its EDGAR time, so we also
time releases by the first large price move.

\section{Data}\label{sec:data}

\subsection{Releases}

The sample contains the earnings releases of S\&P 1500 members from December 2016 to September
2026 that are made before the open or after the close. A release is the first 8-K filing with Item
2.02 by a firm in a quarter, timed by its EDGAR acceptance time. We exclude releases during the
regular session and releases from 16:00 to 16:05 ET. In \SI{6.4}{\percent} of the firm-quarters in
the main sample, the firm files more than one Item 2.02, so the first filing may report a
preliminary result; check 14 in Table~\ref{tab:robust2} excludes these firm-quarters. The 213
releases accepted after 20:00 remain with the releases after the close, although their after-hours
window ends before the filing; check 9 excludes them.

\subsection{Prices}

Prices are 1-minute bars from LSE Data \citep{lse-data}, which cover 04:00 to 20:00 ET on every
trading day, and 00:00 to 04:00 from about September 2022 to September 2026. The data contain no
trades from 20:00 to midnight. LSE Data records only part of the market: at the median, in every
year, its volume from 09:30 to 16:05 is \SI{44}{\percent} of the daily volume reported by Yahoo
Finance (Table~\ref{tab:activity}). We therefore use the data for prices and treat volumes as
shares of the recorded volume. Measured in this way, the coverage of the regular session is
constant over the decade, but the coverage of the pre-market is not. On the morning of the base
session, before any news of the event, \SI{15}{\percent} of events had ten or more pre-market bars
in 2017, against \SI{31}{\percent} in 2026.

Because pre-market coverage changes over time, we also use prices that do not depend on it: the
official open and close of each stock and of SPY from Yahoo Finance's daily data. For
\SI{88}{\percent} of events, the official open is within \SI{0.5}{\percent} of LSE Data's first
trade from 09:30, and the two prices give the same shares in every year
(Appendix~\ref{app:validation}). Of the events in the main sample, 154 have no official open and
are excluded from the tests that use it.

LSE Data includes only firms that still trade, so firms delisted before 2026 are missing. The loss
is largest in the early years: \SI{20}{\percent} of the 2017 releases belong to firms missing from
the data, against \SI{2}{\percent} in 2026 (Table~\ref{tab:activity}).

\subsection{Index membership and controls}

Index membership is taken at each quarter end from the page histories of Wikipedia's lists of
S\&P 500, S\&P 400 and S\&P 600 companies. The S\&P 600 list is usable only from December 2019, so
small caps enter the sample in 2020. Market capitalisation at the prior quarter end is the number
of shares outstanding from SEC XBRL filings times the closing price from Yahoo Finance. VIX data
come from Cboe. We also use Yahoo Finance's daily closes to check LSE Data's prices
(Appendix~\ref{app:validation}).

\subsection{Sample}

Table~\ref{tab:sample} shows the construction of the sample. The largest reduction comes from the
coverage rule: 8{,}737 events have no trade in the data between the EDGAR time and the end of the
release's window. The rule removes \SI{10}{\percent} of S\&P 500 events, \SI{29}{\percent} of
S\&P 400 events and \SI{34}{\percent} of S\&P 600 events, and between 11 and \SI{32}{\percent} of
events depending on the year (Table~\ref{tab:activity}). Because the rule keeps only events with
trading after the release, it selects on the outcome; check 1 removes it. We report the 984
releases of October and November 2026 separately, since LSE Data stops recording overnight trades
in October 2026 (Table~\ref{tab:season}). The main sample contains 29{,}907 events of 1{,}586
firms. December 2016 is included in every regression. Tables and figures by year pool it with
2017, and the H3 regressions exclude it because its 27 events are too few for a separate estimate.

\begin{table}[htbp]
\centering\footnotesize\setlength{\tabcolsep}{3.5pt}
\begin{tabular}{lrrr}
\toprule
& After close & Before open & All \\
\midrule
\multicolumn{4}{l}{\textit{Panel A: sample flow, December 2016 to September 2026}} \\
Item 2.02 filings by index members, first per firm and quarter & & & 48{,}545 \\
\quad released before the open or after 16:05 & & & 44{,}507 \\
\quad firm in LSE Data & & & 39{,}497 \\
\quad LSE Data has bars, reaction session by 30 September 2026 & & & 39{,}359 \\
\quad base and final prices pass the checks & & & 39{,}340 \\
\quad close within 0.5\% of Yahoo Finance & & & 38{,}644 \\
\quad a trade between the release and the end of its window & & & 29{,}907 \\
\midrule
\multicolumn{4}{l}{\textit{Panel B: the main sample}} \\
Events & 15{,}226 & 14{,}681 & 29{,}907 \\
Firms & 1{,}159 & 1{,}061 & 1{,}586 \\
\quad S\&P 500 events & 6{,}034 & 8{,}539 & 14{,}573 \\
\quad S\&P 400 events & 4{,}447 & 3{,}340 & 7{,}787 \\
\quad S\&P 600 events & 4{,}745 & 2{,}802 & 7{,}547 \\
Median $|X|$, per cent & 3.81 & 3.75 & 3.79 \\
Median pre-market bars & 6 & 22 & 13 \\
Mean VIX at the last close & 19.2 & 19.0 & 19.1 \\
\midrule
\multicolumn{4}{l}{\textit{Panel C: October and November 2026, reported separately}} \\
Events, same rules & 454 & 530 & 984 \\
\bottomrule
\end{tabular}
\caption{Sample construction. Panel A applies the sample rules in sequence; each line reports the number of events that remain. Panel B describes the main sample: $X$ is the event return net of SPY, and pre-market bars are LSE Data's 1-minute bars from 04:00 to 09:30 on the reaction morning. Panel C contains the releases after 30 September 2026; LSE Data stops recording overnight trades in October 2026 (Table~\ref{tab:season}).}
\label{tab:sample}
\end{table}

\begin{table}[htbp]
\centering\footnotesize\setlength{\tabcolsep}{3pt}
\begin{tabular}{lrrrrrrrrrr}
\toprule
& \multicolumn{2}{c}{Sample} & \multicolumn{2}{c}{Coverage} & \multicolumn{2}{c}{After hours} & \multicolumn{2}{c}{Shares per trade} & \multicolumn{2}{c}{Fresh price} \\
\cmidrule(lr){2-3}\cmidrule(lr){4-5}\cmidrule(lr){6-7}\cmidrule(lr){8-9}\cmidrule(lr){10-11}
& Releases & No trade & Share of & 10+ bars & Share of & Share of & After & & & \\
Year & lost & after release & volume & before news & volume & trades & hours & Regular & 20:00 & 09:30 \\
\midrule
2017 & 20 & 28 & 44 & 15 & 9.6 & 4.2 & 996 & 405 & 39 & 63 \\
2018 & 17 & 22 & 44 & 18 & 9.1 & 4.8 & 661 & 328 & 32 & 56 \\
2019 & 14 & 21 & 44 & 19 & 8.4 & 5.0 & 475 & 268 & 36 & 57 \\
2020 & 17 & 28 & 44 & 27 & 7.7 & 5.4 & 296 & 201 & 33 & 59 \\
2021 & 15 & 27 & 44 & 24 & 6.8 & 5.0 & 193 & 141 & 30 & 57 \\
2022 & 12 & 32 & 44 & 22 & 7.5 & 5.9 & 159 & 123 & 31 & 53 \\
2023 & 10 & 25 & 44 & 20 & 8.0 & 5.5 & 191 & 127 & 31 & 55 \\
2024 & 7 & 21 & 44 & 22 & 8.8 & 6.5 & 138 & 101 & 33 & 58 \\
2025 & 4 & 14 & 44 & 27 & 8.0 & 6.9 & 94 & 80 & 34 & 66 \\
2026 & 2 & 11 & 44 & 31 & 8.9 & 7.8 & 85 & 74 & 36 & 67 \\
\bottomrule
\end{tabular}
\caption{Coverage, survivorship and after-hours trading in the data, by year, in per cent except for shares per trade. \textit{Sample}: releases lost is the share of releases before the open or after 16:05 whose firm is missing from LSE Data, mostly firms delisted or renamed before 2026; no trade after release is the share of events, after the price checks, that the coverage rule removes. \textit{Coverage}: the median ratio of LSE Data's volume from 09:30 to 16:05 on the base session to Yahoo Finance's daily volume, and the share of main-sample events with ten or more 1-minute bars from 04:00 to 09:30 on the base session, a morning before the news. \textit{After hours}, releases after the close: the after-hours window's share of the volume and of the trades from 16:05 to the next 16:00 close, summed over events; the bars from 16:00 to 16:05 contain the closing auction and are excluded. \textit{Shares per trade}: volume divided by the number of trades in the after-hours window and in the regular hours of the reaction session, for the same releases. \textit{Fresh price}: share of releases after the close whose last trade before 20:00, or before 09:30, occurs in the last five minutes. December 2016 is pooled with 2017.}
\label{tab:activity}
\end{table}

\section{Methodology}\label{sec:method}

\subsection{Event return and windows}

The event return $X$ of each release runs between two closes. It starts at the last trade before
16:00 on the session before the release and ends at the last trade before 16:00 on the first
session after it. We split $X$ into four windows: after hours (16:00--20:00), overnight
(20:00--04:00), pre-market (04:00--09:30) and the regular session (09:30--16:00). The return $x$ of
a window is the log change from the last trade before its start to the last trade before its end,
net of SPY's return over the same window. A window without trades has a zero return before SPY's
return is subtracted, so the four window returns sum to $X$. Checks 17 and 18 replace SPY's return
with beta times SPY's return, or with no benchmark.

A release after the close falls in the after-hours window. A release before the open falls in the
pre-market window, and the evening window before it accounts for almost none of the return. The
data contain overnight trades for few events (none before 2022 and \SI{28}{\percent} of events in
2026), so an overnight move is mostly recorded in the pre-market window. For this reason we combine
the two windows. For releases after the close we call the result the \emph{next morning}
(20:00--09:30); for releases before the open we call the same window the \emph{pre-market}. Our
results do not measure overnight trading as such.

The research plan ends this window at the data's last trade before 09:30. Where the data contain
few pre-market trades, this trade may be hours old, and the move since then is assigned to the
regular session. To avoid this problem, we also end the window at the official open. In this version, the event
return runs between two official closes, the next morning runs from the data's 20:00 price to the
official open, the regular session runs from the official open to the official close, and SPY's
returns come from its official prices. We call the share of the event return from the base close to
the official open the \emph{share priced by the open}. It uses no intraday data, so neither the
coverage of the data nor the coverage rule can affect it. Table~\ref{tab:tests} reports the tests
of the plan, and Table~\ref{tab:official} the version based on the official open.

\subsection{Weighted price contribution}

The weighted price contribution (WPC) of window $w$ is the share of the event return earned in
that window, with larger moves receiving more weight \citep{barclay-warner-1993,barclay-hendershott-2003}:
\begin{equation}
\mathrm{WPC}_w = \sum_i \frac{|X_i|}{\sum_j |X_j|} \cdot \frac{x_{i,w}}{X_i}.
\label{eq:wpc}
\end{equation}
Confidence intervals for the WPC are \SI{95}{\percent} bootstrap percentile intervals from
2{,}000 resamples of release dates, so that events that share a day's market move are drawn
together.

\subsection{Tests}

H1 is a weighted regression of each event's share $s_{i,w} = x_{i,w}/X_i$ on time:
\begin{equation}
s_{i,w} = \alpha_{f(i)} + \beta\, t_i + \gamma_1 D^{\mathrm{2022}}_i + \gamma_2 D^{\mathrm{RH}}_i
+ \gamma_3 D^{\mathrm{BO}}_i + \delta_1 |X_i| + \delta_2 \mathrm{VIX}_i + \theta_{d(i)} + \varepsilon_{i,w},
\label{eq:h1}
\end{equation}
with weights $|X_i|$, so that a regression on a constant alone returns the WPC of
equation~\eqref{eq:wpc}. Here $t_i$ is the release date in years since December 2016,
$\alpha_{f(i)}$ a firm fixed effect, VIX the index level at the last close before the release and
$\theta_{d(i)}$ a day-of-week dummy. Three dummies mark changes in overnight trading:
$D^{\mathrm{2022}}$ from September 2022, when LSE Data starts to store overnight trades;
$D^{\mathrm{RH}}$ for the second and third quarters of 2023, around the launch of Robinhood's
overnight trading on 16 May 2023; and $D^{\mathrm{BO}}$ for the third quarter of 2024, which
contains the Blue Ocean outage of 5 August. Years and quarters in the tables refer to the reaction
session. H1 predicts $\beta > 0$ for the after-hours share of releases after the close.

With $D^{\mathrm{2022}}$ in the model, $\beta$ measures the slope within the periods before and
after September 2022, not the change over the decade; $t_i$ and $D^{\mathrm{2022}}$ have a
correlation of 0.85. We therefore also report the implied change from December 2016 to September
2026, $9.83\,\hat\beta + \hat\gamma_1$, and a model without the event dummies that replaces
$\beta\,t_i$ with dummies for 2020--2023 and 2024--2026. The coefficient on the second dummy is the
change from 2016--2019 to 2024--2026. Because the shares fall until 2021 and then recover, a linear
trend fits them poorly, and we give more weight to the period model than to the trend.

H2 adds the index tier and its interaction with $t_i$; a check uses market-capitalisation terciles
within each year instead. H2 predicts a negative interaction for the S\&P 600.

H3 asks how much of the event return the price at a given time already reflects
\citep{biais-1999-preopening,barclay-hendershott-2003}. For each year and release timing, we
estimate
\begin{equation}
X_i = a + b\, r_i^{(\tau)} + e_i,
\label{eq:h3}
\end{equation}
where $r_i^{(\tau)}$ is the return from the last close before the release to time $\tau$. H3
predicts that the slope at $\tau = $~09:30 moves toward 1 and that its $R^2$ rises over the years.
If prices follow a martingale, the slope is 1 regardless of how much of the news is priced by
$\tau$, so only the $R^2$ is informative \citep{vanbommel-2011}. We also report an $R^2$ with both
returns winsorised at the year's 1st and 99th percentiles, an event-level test of the trend in the
slope, and the $R^2$ with the official open in place of the last trade before 09:30.

\subsection{Inference and power}

Standard errors in equation~\eqref{eq:h1} are clustered by firm and by release date
\citep{petersen-2009,cameron-2011}. Firms with a single event in a regression drop out, since the
firm fixed effect leaves no variation for them. The trend varies only over time, and releases
cluster in earnings seasons, so shocks common to a season can make date clusters too narrow. For
this reason, we also cluster by firm and calendar quarter, and we estimate the trend in the quarterly WPC
with Newey--West standard errors \citep{newey-west-1987}. The trend in the yearly values of
equation~\eqref{eq:h3} is an OLS slope over the ten years 2017--2026.

For each test we report the minimum detectable effect. At the \SI{5}{\percent} level with
\SI{80}{\percent} power, it is about 2.8 standard errors.

\section{Results}\label{sec:results}

\subsection{Window shares by year}

\begin{figure}[htbp]\centering
\includegraphics[width=\linewidth]{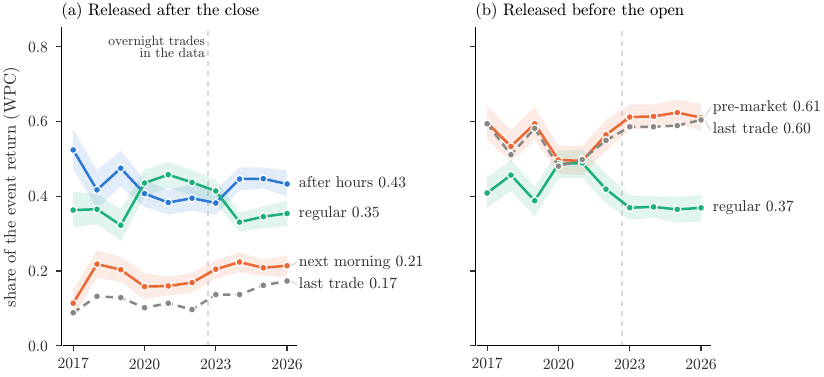}
\caption{Each window's share of the event return (WPC, equation~\ref{eq:wpc}) by year, with
\SI{95}{\percent} bootstrap intervals over release dates as shaded bands. Solid lines end the
extended hours at the official open: the next morning in panel~(a) and the pre-market in
panel~(b) run from 20:00 to the official open, and the regular session runs from the official open
to the close. The gray dashed line shows the 20:00--09:30 window measured to the data's last trade
before 09:30, as defined in the research plan. Panel~(a): releases after the close. Panel~(b):
releases before the open. The dashed vertical line marks September 2022, from which the data
contain some overnight trades; the 2022 point covers the whole year. December 2016 is pooled with
2017. Data: London Strategic Edge; official prices from Yahoo Finance.}
\label{fig:wpc}
\end{figure}

Figure~\ref{fig:wpc} shows each window's share of the event return by year, with the extended
hours ending at the official open. For releases after the close, over the whole period, the
after-hours session accounts for 0.42 of the move, the next morning for 0.20 and the regular
session for 0.38. For releases before the open, the pre-market accounts for 0.58 and the regular
session for 0.40. With the 09:30 price of the plan, the data's last trade before 09:30, the next
morning accounts for 0.13 and the regular session for 0.44 after the close (dashed line in
panel~a).

Part of the after-hours low in 2020--2023 reflects the composition of the sample. S\&P 600 firms,
whose after-hours share is lower, enter in 2020 and account for a third or more of the releases
after the close from then on. For S\&P 500 and 400 firms alone, the after-hours share is 0.42 to
0.43 in 2022 and 2023, against 0.38 to 0.39 for the full sample. The same years also have the most
events without trades after the release (Table~\ref{tab:activity}) and most of the events whose
close disagrees with Yahoo Finance (Appendix~\ref{app:validation}). Composition and data quality do
not explain the whole decline, however. The share priced by the official open, which requires no
intraday data, also falls in 2020--2023 for S\&P 500 and 400 firms, from 0.65 to 0.59 after the
close.

\subsection{H1: no trend in the after-hours share}

For releases after the close, the after-hours session accounted for 0.52 of the move in 2017, 0.38
to 0.41 in 2020--2023 and 0.43 in 2026. With controls and firm fixed effects, the trend is
$+0.0007$ per year with a standard error of 0.0052 (Table~\ref{tab:tests}, Panel~A), so the planned
test does not support H1. The implied change over the decade is $+0.037$ (0.033), with a
\SI{95}{\percent} confidence interval from $-0.027$ to $+0.101$. The minimum detectable effect is
about 0.015 per year for the trend alone, or 0.14 over the decade, and about 0.09 for the implied
change. We can reject neither zero nor an increase of 0.10, a quarter of the level.

The share does not follow a linear path. Relative to 2016--2019, it is 0.037 lower in 2020--2023
(standard error 0.020) and 0.031 higher in 2024--2026 (0.020). The alternative specifications
agree on a small positive change but differ on its significance (Tables~\ref{tab:robust}
and~\ref{tab:robust2}, column~1). The change from 2016--2019 to 2024--2026 is not significant in
the main sample, in the constant panel or with clustering by quarter. It is $+0.045$ (0.018)
without the coverage rule, $+0.040$ (0.021) net of beta times SPY's return and $+0.042$ (0.020)
with raw returns (checks 1, 17 and 18). Under the last two benchmarks, the \SI{95}{\percent}
interval of the implied change ends at 0.12 and 0.13, respectively. The quarterly series has a
trend of $-0.0050$ (0.0041), and among stocks with dense pre-market coverage before the news the
estimated change in the share is negative. H1 predicted an increase in line with the growth of
extended-hours trading. The data instead show a decline until 2021 and a return to the earlier
level or slightly above it.

Part of the remaining increase is not a response to earnings. The after-hours window starts at the
last trade before 16:00, so it includes the move to the closing auction, which accounts for 0.03 of
the event return. When the window starts at 16:05, the change from 2016--2019 to 2024--2026 is
$+0.013$ (0.021) (check 21). Of the remaining after-hours share, half occurs before the EDGAR
acceptance time (0.20 of the event return) and half after it (0.19), which indicates how far the
filing lags the newswire release.

After-hours volume did not increase either, which may explain why the share did not. For releases
after the close, the after-hours share of the volume from 16:05 to the next close is
\SI{9.6}{\percent} in 2017, \SI{6.8}{\percent} in 2021 and \SI{8.9}{\percent} in 2026; it falls and
recovers in the same way as the after-hours WPC (Table~\ref{tab:activity}). Its share of trades
rises from 4.2 to \SI{7.8}{\percent}. Trade sizes declined in the regular session as well, from 405
to 74 shares per trade, in line with the market as a whole \citep{ohara-2014-oddlots}, but they
declined more after hours, from 996 to 85 shares. An after-hours trade was two and a half times the
size of a regular-session trade in 2017 and is about the same size in 2026, which is consistent
with a larger role for retail orders \citep{boehmer-2021-retail}.

The two event dummies in the plan are of limited use. Each covers whole quarters and thus
includes releases from before the event: \SI{85}{\percent} of the releases in the second quarter of
2023 react before 16 May, and \SI{64}{\percent} of those in the third quarter of 2024 react before
5 August. With the dummies dated by reaction session (check 20), the after-hours share is 0.058
lower (0.025) in the six months after Robinhood's launch and 0.069 higher (0.028) in the two weeks
of the Blue Ocean outage. The second estimate rests on 137 releases that also fall in the market
turmoil of early August 2024. We do not interpret either estimate as an effect of overnight
trading. Among the controls, larger moves place more of the response in the extended windows, and
a higher VIX places more of it in the regular session.

\begin{table}[htbp]
\centering\footnotesize\setlength{\tabcolsep}{3.5pt}
\begin{tabular}{lccc}
\toprule
& (1) & (2) & (3) \\
& After close & After close & Before open \\
& 16:00--20:00 & 20:00--09:30 & 20:00--09:30 \\
\midrule
\multicolumn{4}{l}{\textit{Panel A: H1, the trend}} \\
Trend, per year & 0.0007 & 0.0090$^{**}$ & 0.0085$^{*}$ \\
 & (0.0052) & (0.0038) & (0.0049) \\
From September 2022 & 0.0299 & $-$0.0102 & 0.0170 \\
 & (0.0262) & (0.0185) & (0.0287) \\
Two quarters after Robinhood's overnight launch & $-$0.0518$^{*}$ & 0.0126 & 0.0027 \\
 & (0.0264) & (0.0186) & (0.0311) \\
Quarter of the Blue Ocean outage & $-$0.0260 & 0.0362$^{**}$ & 0.0642$^{**}$ \\
 & (0.0299) & (0.0184) & (0.0270) \\
$|X|$ & 0.2011$^{**}$ & 0.1889$^{***}$ & 0.1640 \\
 & (0.0858) & (0.0697) & (0.1016) \\
VIX at the last close & $-$0.0026$^{**}$ & $-$0.0014 & $-$0.0045$^{***}$ \\
 & (0.0010) & (0.0009) & (0.0012) \\
Change, December 2016 to September 2026 & 0.037 & 0.078 & 0.101 \\
 & (0.033) & (0.024) & (0.029) \\
\quad 95\% interval & [$-$0.027, 0.101] & [0.030, 0.125] & [0.045, 0.157] \\
2020--2023 against 2016--2019 & $-$0.0367$^{*}$ & 0.0012 & 0.0235 \\
 & (0.0195) & (0.0143) & (0.0211) \\
2024--2026 against 2016--2019 & 0.0305 & 0.0358$^{**}$ & 0.0683$^{***}$ \\
 & (0.0200) & (0.0149) & (0.0203) \\
Events & 15{,}060 & 15{,}060 & 14{,}502 \\
Within $R^2$ & 0.000 & 0.001 & 0.001 \\
\midrule
\multicolumn{4}{l}{\textit{Panel B: H2, the trend by index tier and by size}} \\
Trend, S\&P 500 & $-$0.0036 & 0.0093$^{**}$ & 0.0076 \\
 & (0.0057) & (0.0039) & (0.0051) \\
\quad $\times$ S\&P 400 & 0.0140$^{**}$ & $-$0.0012 & 0.0050 \\
 & (0.0064) & (0.0044) & (0.0067) \\
\quad $\times$ S\&P 600 & $-$0.0018 & 0.0044 & 0.0097 \\
 & (0.0085) & (0.0056) & (0.0096) \\
\quad $\times$ S\&P 600, reaction sessions from 2020 & $-$0.0014 & $-$0.0013 & 0.0038 \\
 & (0.0099) & (0.0065) & (0.0104) \\
Events & 15{,}060 & 15{,}060 & 14{,}502 \\
Trend, top size tercile & $-$0.0034 & 0.0119$^{***}$ & 0.0069 \\
 & (0.0064) & (0.0045) & (0.0059) \\
\quad $\times$ middle tercile & 0.0126$^{**}$ & $-$0.0038 & $-$0.0018 \\
 & (0.0063) & (0.0044) & (0.0061) \\
\quad $\times$ bottom tercile & 0.0030 & $-$0.0036 & 0.0007 \\
 & (0.0067) & (0.0044) & (0.0069) \\
\quad $\times$ bottom tercile, reaction sessions from 2020 & 0.0047 & $-$0.0021 & $-$0.0052 \\
 & (0.0096) & (0.0067) & (0.0103) \\
Events & 14{,}872 & 14{,}872 & 14{,}241 \\
\bottomrule
\end{tabular}
\caption{Tests of H1 and H2. The dependent variable is a window's share of the event return, $s = x/X$, with both returns net of SPY. Weighted least squares with weights $|X|$, firm fixed effects, day-of-week dummies and, in every column, the controls of Panel A. The trend is measured in years since December 2016. Panel B adds dummies for the index tier, or for the size tercile within each year, and their interactions with the trend; the S\&P 500 and the top tercile are the base groups. The lines marked from 2020 re-estimate the models on reaction sessions from 2020, when the S\&P 600 enters, so that all groups cover the same years and the terciles the same universe. The change from December 2016 to September 2026 is the trend times 9.83 years plus the September 2022 dummy. The two period lines come from a model that replaces the trend and the event dummies with dummies for 2020--2023 and 2024--2026. The event dummies cover whole quarters and therefore include releases from before each event; check 20 of Table~\ref{tab:robust2} dates them by day. Standard errors, clustered by firm and by release date, are in parentheses. $^{*}$, $^{**}$ and $^{***}$ denote significance at the 10\%, 5\% and 1\% levels. Firms with a single event in a column drop out.}
\label{tab:tests}
\end{table}

\subsection{The next morning and stale pre-market prices}

Under the plan's definition, the next-morning share increases. For releases after the close, it
rises from 0.09 in 2017 to 0.17 in 2026, while the regular session's share falls. With the event
dummies of the plan, the trend is 0.0090 per year (0.0038), and the implied change over the decade
is $+0.078$ (0.024) (Table~\ref{tab:tests}). The increase passes the checks that rely on the bars
themselves. It holds in a constant panel of firms (check 6b), for events whose 20:00 and 09:30
prices both come from the last five minutes (check 8), and among stocks with ten or more
pre-market bars on the morning before the news, a measure of coverage that the news cannot affect
(check 13). In the last group, the change from 2016--2019 to 2024--2026 is $+0.079$ (0.026).

The increase does not survive a comparison with official prices. The plan ends the morning at the
data's last trade before 09:30. If that trade is stale, the move between it and the open is
assigned to the regular session, and the data contain fewer pre-market trades in the early years
(Table~\ref{tab:activity}). Table~\ref{tab:official} instead ends the morning at the official open.
The next-morning share is then 0.11 in 2017, 0.20 to 0.22 in 2018 and 2019, 0.16 to 0.17 in
2020--2022 and 0.21 to 0.22 in 2024--2026 (Panel~A). The difference between the two definitions is
the move from the last pre-market trade in the data to the open. It amounts to 0.07 to 0.09 of the
event return in 2018, 2019 and 2024, and to 0.04 to 0.05 in 2025 and 2026, when the data contain
more pre-market trades. Half of the increase under the plan's definition reflects the narrowing of
this difference.

Measured to the official open, the next-morning share in 2024--2026 exceeds that of 2016--2019 by
0.016 (0.017), and by 0.006 (0.016) without the coverage rule (Table~\ref{tab:official}, Panel~B,
column~2). The increase remains only among stocks with dense pre-market trading before the news
($+0.082$, standard error 0.025), a selected fifth of the sample whose composition changes over
time. We conclude that the bars overstate the shift to the next morning and that, for the average
release after the close, the shift is small.

For releases before the open, the same comparison supports a result that the first draft of this
paper questioned. For these releases, the pre-market share measured to the last trade rises by
0.068 (0.020) from 2016--2019 to 2024--2026, but not among well-covered stocks (checks 5 and 13),
so changes in coverage could explain it. The official prices do not support this explanation: the
share from the base close to the official open rises by 0.072 (0.019), by 0.098 (0.017) without the
coverage rule and by 0.055 (0.020) in the constant panel (column~3). Among stocks with ten or more
pre-market bars before the news, the estimate is 0.008 with a standard error of 0.040, too
imprecise to contradict the other estimates.

\subsection{The share priced by the open}

The share of the event return from the base close to the official open summarises the extended
hours and requires no intraday data. For releases after the close, it is 0.65 in 2017--2019, 0.57
in 2020--2023 and 0.66 in 2024--2026; for S\&P 500 and 400 firms alone, it is 0.65, 0.59 and 0.68.
For releases before the open, it is 0.58, 0.57 and 0.63.

For releases after the close, the main pattern is the decline and recovery. Within firms and with
controls, the share priced by the open is 0.045 lower in 2020--2023 than in 2016--2019 (standard
error 0.022) and 0.049 higher in 2024--2026 (0.022) (Table~\ref{tab:official}, Panel~B). The second
estimate is 0.054 without the coverage rule and 0.057 for S\&P 500 and 400 firms, and it loses
significance in the constant panel (0.044, standard error 0.027). The evening and the next morning
each contribute a part that is too small to be significant on its own: $+0.031$ (0.020) and
$+0.016$ (0.017). For releases before the open, there is no decline within firms, and the share is
0.072 higher in 2024--2026 (0.019). LSE Data's own first trade from 09:30 gives the same results as
the official open (last two rows). The regular session's share therefore fell by about 0.05 after
the close, against 0.115 (0.035) under the plan's definition, and by 0.05 to 0.10 before the open.

To assess whether this change is specific to earnings, the last column of Panel~A reports the same
share for the same firms on days without earnings news: the part of a day's return, net of SPY,
that is priced by the open, on days with a move of \SI{1}{\percent} or more. It is 0.16 in
2017--2019, 0.17 in 2020--2023 and 0.20 in 2024--2026. On ordinary days, too, the open thus prices
about 0.04 more of the day's move than before, and the increase on earnings days is of similar
size. What distinguishes earnings days is the decline of 2020--2023 after the close, which ordinary
days do not show. These are the years of the pandemic and of the highest retail trading volumes;
we have no test of why the regular session accounted for more of the earnings response in this
period.

Estimates split by the sign of the news are imprecise (checks 19a and 19b). Under the plan's
definition, the morning shares rise more after good news than after bad news, and the change in the
after-hours share is the same for both signs, 0.03 and not significant.

With clustering by firm and quarter, the morning trends of the plan are significant only at the
\SI{10}{\percent} level or not at all (check 11). Before September 2022, no window has a trend
(checks 7a and 7b). From then on, the after-hours share rises by 0.016 per year (0.008) and the
next-morning share by 0.023 (0.006), which reflects the recovery from the 2020--2023 low.

\begin{table}[htbp]
\centering\scriptsize\setlength{\tabcolsep}{3pt}
\begin{tabular}{lrrrrrrrrrr}
\toprule
& \multicolumn{5}{c}{After close} & \multicolumn{4}{c}{Before open} & No news \\
\cmidrule(lr){2-6}\cmidrule(lr){7-10}\cmidrule(lr){11-11}
& & After & \multicolumn{2}{c}{Next morning, to} & Regular, & & \multicolumn{2}{c}{Pre-market, to} & Regular, & Priced \\
\cmidrule(lr){4-5}\cmidrule(lr){8-9}
Year & Events & hours & last trade & open & from open & Events & last trade & open & from open & by open \\
\midrule
\multicolumn{11}{l}{\textit{Panel A: WPC by year}} \\
2017 & 749 & 0.52 & 0.09 & 0.11 & 0.36 & 920 & 0.59 & 0.60 & 0.41 & 0.16 \\
2018 & 914 & 0.42 & 0.13 & 0.22 & 0.37 & 992 & 0.51 & 0.53 & 0.46 & 0.15 \\
2019 & 917 & 0.48 & 0.13 & 0.20 & 0.32 & 1{,}089 & 0.58 & 0.59 & 0.39 & 0.17 \\
2020 & 1{,}449 & 0.41 & 0.11 & 0.16 & 0.44 & 1{,}325 & 0.48 & 0.50 & 0.49 & 0.18 \\
2021 & 1{,}618 & 0.38 & 0.11 & 0.16 & 0.46 & 1{,}497 & 0.50 & 0.49 & 0.49 & 0.15 \\
2022 & 1{,}612 & 0.39 & 0.10 & 0.17 & 0.44 & 1{,}417 & 0.55 & 0.56 & 0.42 & 0.17 \\
2023 & 1{,}832 & 0.38 & 0.14 & 0.20 & 0.41 & 1{,}640 & 0.58 & 0.61 & 0.37 & 0.17 \\
2024 & 2{,}001 & 0.44 & 0.14 & 0.22 & 0.33 & 1{,}857 & 0.58 & 0.61 & 0.37 & 0.20 \\
2025 & 2{,}146 & 0.45 & 0.16 & 0.21 & 0.35 & 2{,}171 & 0.59 & 0.62 & 0.36 & 0.19 \\
2026 & 1{,}786 & 0.43 & 0.17 & 0.21 & 0.35 & 1{,}660 & 0.61 & 0.61 & 0.37 & 0.22 \\
\bottomrule
\end{tabular}

\vspace{6pt}
\begin{tabular}{lrrrrrr}
\toprule
\multicolumn{7}{l}{\textit{Panel B: the share priced by the official open against 2016--2019}} \\
& \multicolumn{2}{c}{(1) After close, close--open} & \multicolumn{2}{c}{(2) After close, 20:00--open} & \multicolumn{2}{c}{(3) Before open, close--open} \\
\cmidrule(lr){2-3}\cmidrule(lr){4-5}\cmidrule(lr){6-7}
& 2020--23 & 2024--26 & 2020--23 & 2024--26 & 2020--23 & 2024--26 \\
\midrule
Main sample & $-$0.0450$^{**}$ & 0.0494$^{**}$ & $-$0.0084 & 0.0163 & 0.0134 & 0.0717$^{***}$ \\
\quad $N$ = 14{,}967; 14{,}443 & (0.0215) & (0.0217) & (0.0164) & (0.0169) & (0.0191) & (0.0186) \\[1pt]
No coverage rule & $-$0.0420$^{**}$ & 0.0541$^{***}$ & $-$0.0177 & 0.0064 & 0.0237 & 0.0980$^{***}$ \\
\quad $N$ = 19{,}012; 19{,}053 & (0.0188) & (0.0192) & (0.0157) & (0.0161) & (0.0172) & (0.0165) \\[1pt]
S\&P 500 and 400 & $-$0.0354$^{*}$ & 0.0565$^{***}$ & $-$0.0045 & 0.0168 & $-$0.0098 & 0.0504$^{***}$ \\
\quad $N$ = 10{,}347; 11{,}760 & (0.0210) & (0.0218) & (0.0163) & (0.0173) & (0.0190) & (0.0181) \\[1pt]
Constant panel & $-$0.0429$^{*}$ & 0.0436 & $-$0.0101 & 0.0267 & $-$0.0144 & 0.0550$^{***}$ \\
\quad $N$ = 6{,}213; 8{,}988 & (0.0259) & (0.0271) & (0.0201) & (0.0217) & (0.0202) & (0.0196) \\[1pt]
10+ bars before the news & $-$0.0191 & 0.0326 & 0.0268 & 0.0820$^{***}$ & $-$0.0383 & 0.0080 \\
\quad $N$ = 3{,}309; 3{,}313 & (0.0419) & (0.0416) & (0.0220) & (0.0252) & (0.0441) & (0.0397) \\[1pt]
Firm and quarter clusters & $-$0.0450$^{*}$ & 0.0494$^{**}$ & $-$0.0084 & 0.0163 & 0.0134 & 0.0717$^{***}$ \\
\quad $N$ = 14{,}967; 14{,}443 & (0.0267) & (0.0237) & (0.0219) & (0.0230) & (0.0242) & (0.0250) \\[1pt]
LSE Data's open & $-$0.0418$^{**}$ & 0.0494$^{**}$ & $-$0.0052 & 0.0207 & 0.0089 & 0.0605$^{***}$ \\
\quad $N$ = 14{,}949; 14{,}452 & (0.0212) & (0.0216) & (0.0162) & (0.0165) & (0.0195) & (0.0190) \\[1pt]
LSE Data's open, no coverage rule & $-$0.0430$^{**}$ & 0.0489$^{**}$ & $-$0.0183 & 0.0058 & 0.0227 & 0.0886$^{***}$ \\
\quad $N$ = 18{,}932; 18{,}997 & (0.0188) & (0.0192) & (0.0156) & (0.0160) & (0.0178) & (0.0171) \\[1pt]
\bottomrule
\end{tabular}
\caption{The official open. Here the extended hours end at the official opening price from Yahoo Finance's daily data instead of LSE Data's last trade before 09:30, and the event return runs between two official closes, net of SPY's official prices. Panel A: WPC by year in the main sample, for events with an official open; \textit{last trade} is the plan's 20:00--09:30 window, \textit{open} runs from LSE Data's 20:00 price to the official open, and \textit{regular, from open} runs from the official open to the official close. \textit{No news}: for the same firms and years, the WPC of the close-to-open return within the close-to-close return, on days at least five sessions away from the firm's reaction sessions with an absolute return of 1\% or more, net of SPY and based on official prices. December 2016 is pooled with 2017. Panel B: weighted least squares as in Table~\ref{tab:tests}, with weights and the $|X|$ control from official closes, firm fixed effects, and dummies for reaction sessions in 2020--2023 and in 2024--2026 instead of the trend and the event dummies; 2016--2019 is the base. Columns (1) and (3): the share of the event return from the base close to the official open. Column (2): the same share less LSE Data's after-hours return. The last two rows replace the official open with LSE Data's own first trade from 09:30, and the official closes with LSE Data's closes. Standard errors, clustered by firm and release date unless stated otherwise, are in parentheses. Under each label, $N$ gives the numbers of post-close and pre-open events.}
\label{tab:official}
\end{table}

\begin{table}[htbp]
\centering\scriptsize\setlength{\tabcolsep}{3pt}
\begin{tabular}{lrrrrrr}
\toprule
& \multicolumn{2}{c}{(1) After close, 16:00--20:00} & \multicolumn{2}{c}{(2) After close, 20:00--09:30} & \multicolumn{2}{c}{(3) Before open, 20:00--09:30} \\
\cmidrule(lr){2-3}\cmidrule(lr){4-5}\cmidrule(lr){6-7}
& Trend & Change & Trend & Change & Trend & Change \\
\midrule
Main specification & 0.0007 & 0.0305 & 0.0090$^{**}$ & 0.0358$^{**}$ & 0.0085$^{*}$ & 0.0683$^{***}$ \\
\quad $N$ = 15{,}060; 14{,}502 & (0.0052) & (0.0200) & (0.0038) & (0.0149) & (0.0049) & (0.0203) \\[1pt]
1. No coverage rule & 0.0052 & 0.0451$^{**}$ & 0.0091$^{***}$ & 0.0324$^{**}$ & 0.0204$^{***}$ & 0.1367$^{***}$ \\
\quad $N$ = 19{,}151; 19{,}152 & (0.0044) & (0.0181) & (0.0035) & (0.0138) & (0.0042) & (0.0190) \\[1pt]
2. Price-based release time & 0.0002 & 0.0287 & 0.0090$^{**}$ & 0.0394$^{**}$ & 0.0085$^{*}$ & 0.0683$^{***}$ \\
\quad $N$ = 14{,}377; 14{,}502 & (0.0051) & (0.0204) & (0.0040) & (0.0155) & (0.0049) & (0.0203) \\[1pt]
3. $X$ to the next close & 0.0062 & 0.0246 & 0.0007 & 0.0109 & 0.0013 & 0.0442$^{**}$ \\
\quad $N$ = 14{,}967; 14{,}443 & (0.0048) & (0.0208) & (0.0037) & (0.0143) & (0.0051) & (0.0208) \\[1pt]
4. Equal weights, $|X| \geq 1\%$ & 0.0002 & 0.0549 & 0.0080 & 0.0547$^{**}$ & 0.0055 & 0.0752$^{***}$ \\
\quad $N$ = 12{,}766; 12{,}246 & (0.0085) & (0.0350) & (0.0059) & (0.0213) & (0.0071) & (0.0289) \\[1pt]
5. 30+ pre-market bars & $-$0.0186$^{*}$ & $-$0.0529 & 0.0050 & 0.0420$^{*}$ & $-$0.0013 & 0.0211 \\
\quad $N$ = 3{,}930; 6{,}272 & (0.0096) & (0.0372) & (0.0061) & (0.0242) & (0.0067) & (0.0269) \\[1pt]
6a. S\&P 500 and 400 & $-$0.0007 & 0.0370$^{*}$ & 0.0050 & 0.0505$^{***}$ & 0.0014 & 0.0484$^{**}$ \\
\quad $N$ = 10{,}371; 11{,}771 & (0.0054) & (0.0214) & (0.0041) & (0.0156) & (0.0047) & (0.0203) \\[1pt]
6b. Constant panel & $-$0.0033 & 0.0118 & 0.0011 & 0.0524$^{***}$ & 0.0036 & 0.0598$^{***}$ \\
\quad $N$ = 6{,}213; 8{,}988 & (0.0074) & (0.0257) & (0.0056) & (0.0199) & (0.0055) & (0.0221) \\[1pt]
6c. Ticker unchanged & 0.0012 & 0.0345$^{*}$ & 0.0085$^{**}$ & 0.0286$^{*}$ & 0.0085$^{*}$ & 0.0751$^{***}$ \\
\quad $N$ = 14{,}427; 14{,}075 & (0.0053) & (0.0207) & (0.0040) & (0.0162) & (0.0049) & (0.0203) \\[1pt]
7a. Before September 2022 & $-$0.0118 & -- & $-$0.0015 & -- & 0.0027 & -- \\
\quad $N$ = 6{,}794; 6{,}749 & (0.0078) &  & (0.0056) &  & (0.0073) &  \\[1pt]
7b. From September 2022 & 0.0163$^{**}$ & -- & 0.0234$^{***}$ & -- & 0.0148 & -- \\
\quad $N$ = 8{,}155; 7{,}621 & (0.0082) &  & (0.0065) &  & (0.0093) &  \\[1pt]
\bottomrule
\end{tabular}
\caption{Robustness. \textit{Trend} repeats the H1 regression of Table~\ref{tab:tests}, Panel A, on another sample or with another definition. \textit{Change} is the coefficient on a dummy for reaction sessions in 2024--2026, estimated with a dummy for 2020--2023, 2016--2019 as the base, the same controls and firm fixed effects, and no event dummies; it measures the increase without allowing the September 2022 dummy to absorb it. The checks of the research plan: (1) keeps events with no trade after the release; (2) excludes releases after the close whose first half-way price move comes before 16:05, which a newswire release before the EDGAR time would place in or at the end of the regular session; (3) measures $X$ to the close of the following session; (4) drops the weights and excludes events with $|X|$ below 1\%; (5) keeps events with 30 or more pre-market bars on the reaction morning; (6a) excludes S\&P 600 firms; (6b) keeps the 449 firms with a release in every year from 2017 to 2026; (6c) excludes firms whose ticker changed; (7a) and (7b) split the sample at September 2022. Standard errors as in Table~\ref{tab:tests}. Under each label, $N$ gives the numbers of post-close and pre-open events.}
\label{tab:robust}
\end{table}
\begin{table}[htbp]
\centering\scriptsize\setlength{\tabcolsep}{3pt}
\begin{tabular}{lrrrrrr}
\toprule
& \multicolumn{2}{c}{(1) After close, 16:00--20:00} & \multicolumn{2}{c}{(2) After close, 20:00--09:30} & \multicolumn{2}{c}{(3) Before open, 20:00--09:30} \\
\cmidrule(lr){2-3}\cmidrule(lr){4-5}\cmidrule(lr){6-7}
& Trend & Change & Trend & Change & Trend & Change \\
\midrule
8. Fresh window-end prices & $-$0.0106 & $-$0.0258 & 0.0077 & 0.0487$^{**}$ & 0.0008 & 0.0498$^{**}$ \\
\quad $N$ = 4{,}405; 11{,}440 & (0.0094) & (0.0341) & (0.0054) & (0.0228) & (0.0053) & (0.0212) \\[1pt]
9. No releases after 20:00 & 0.0004 & 0.0312 & 0.0098$^{***}$ & 0.0341$^{**}$ & 0.0085$^{*}$ & 0.0683$^{***}$ \\
\quad $N$ = 14{,}851; 14{,}502 & (0.0052) & (0.0204) & (0.0038) & (0.0151) & (0.0049) & (0.0203) \\[1pt]
10. No event dummies & 0.0051$^{*}$ & -- & 0.0077$^{***}$ & -- & 0.0117$^{***}$ & -- \\
\quad $N$ = 15{,}060; 14{,}502 & (0.0029) &  & (0.0021) &  & (0.0026) &  \\[1pt]
11. Firm and quarter clusters & 0.0007 & 0.0305 & 0.0090$^{*}$ & 0.0358$^{*}$ & 0.0085 & 0.0683$^{**}$ \\
\quad $N$ = 15{,}060; 14{,}502 & (0.0052) & (0.0189) & (0.0053) & (0.0195) & (0.0058) & (0.0261) \\[1pt]
12. From 2020 & 0.0080 & -- & 0.0166$^{***}$ & -- & 0.0163$^{**}$ & -- \\
\quad $N$ = 12{,}468; 11{,}499 & (0.0072) &  & (0.0054) &  & (0.0079) &  \\[1pt]
13. 10+ bars before the news & $-$0.0159 & $-$0.0526 & 0.0078 & 0.0792$^{***}$ & $-$0.0039 & $-$0.0039 \\
\quad $N$ = 3{,}322; 3{,}325 & (0.0116) & (0.0418) & (0.0067) & (0.0259) & (0.0095) & (0.0388) \\[1pt]
14. One Item 2.02 filing & $-$0.0002 & 0.0311 & 0.0089$^{**}$ & 0.0363$^{**}$ & 0.0122$^{**}$ & 0.0848$^{***}$ \\
\quad $N$ = 14{,}129; 13{,}604 & (0.0052) & (0.0198) & (0.0039) & (0.0153) & (0.0050) & (0.0210) \\[1pt]
15. Minutes to window end & 0.0006 & 0.0301 & 0.0096$^{**}$ & 0.0397$^{***}$ & 0.0079 & 0.0636$^{***}$ \\
\quad $N$ = 15{,}060; 14{,}502 & (0.0052) & (0.0200) & (0.0038) & (0.0152) & (0.0049) & (0.0208) \\[1pt]
16. Quarterly series & $-$0.0050 & $-$0.0285 & 0.0062$^{***}$ & 0.0391$^{***}$ & 0.0061 & 0.0302$^{**}$ \\
\quad 39; 39 quarters & (0.0041) & (0.0188) & (0.0023) & (0.0141) & (0.0038) & (0.0147) \\[1pt]
17. Net of beta $\times$ SPY & 0.0039 & 0.0395$^{*}$ & 0.0092$^{**}$ & 0.0330$^{**}$ & 0.0122$^{**}$ & 0.0665$^{***}$ \\
\quad $N$ = 14{,}701; 14{,}062 & (0.0055) & (0.0211) & (0.0040) & (0.0155) & (0.0053) & (0.0210) \\[1pt]
18. No benchmark & 0.0058 & 0.0416$^{**}$ & 0.0054 & 0.0303$^{**}$ & 0.0117$^{**}$ & 0.0781$^{***}$ \\
\quad $N$ = 15{,}056; 14{,}502 & (0.0052) & (0.0204) & (0.0040) & (0.0151) & (0.0049) & (0.0203) \\[1pt]
19a. Good news, $X > 0$ & 0.0133 & 0.0273 & 0.0102$^{*}$ & 0.0805$^{***}$ & 0.0262$^{***}$ & 0.1094$^{***}$ \\
\quad $N$ = 7{,}415; 7{,}231 & (0.0087) & (0.0323) & (0.0061) & (0.0229) & (0.0086) & (0.0311) \\[1pt]
19b. Bad news, $X < 0$ & $-$0.0160$^{**}$ & 0.0302 & 0.0112$^{**}$ & 0.0125 & $-$0.0044 & 0.0548$^{*}$ \\
\quad $N$ = 7{,}495; 7{,}086 & (0.0070) & (0.0304) & (0.0051) & (0.0204) & (0.0076) & (0.0279) \\[1pt]
20. Event dummies by date & 0.0012 & -- & 0.0089$^{**}$ & -- & 0.0089$^{*}$ & -- \\
\quad $N$ = 15{,}060; 14{,}502 & (0.0051) &  & (0.0038) &  & (0.0049) &  \\[1pt]
21. After hours from 16:05 & 0.0011 & 0.0126 & 0.0090$^{**}$ & 0.0358$^{**}$ & 0.0085$^{*}$ & 0.0683$^{***}$ \\
\quad $N$ = 15{,}060; 14{,}502 & (0.0052) & (0.0211) & (0.0038) & (0.0149) & (0.0049) & (0.0203) \\[1pt]
22. No control for $|X|$ & 0.0013 & 0.0357$^{*}$ & 0.0095$^{**}$ & 0.0411$^{***}$ & 0.0086$^{*}$ & 0.0708$^{***}$ \\
\quad $N$ = 15{,}060; 14{,}502 & (0.0052) & (0.0202) & (0.0037) & (0.0145) & (0.0049) & (0.0203) \\[1pt]
\bottomrule
\end{tabular}
\caption{Robustness: checks added after the main results and not included in the research plan. Columns as in Table~\ref{tab:robust}. (8) keeps events whose 20:00 and 09:30 prices come from trades in the last five minutes before those times (for releases before the open, only the 09:30 price); (9) excludes releases after 20:00; (10) drops the three event dummies; (11) clusters by firm and by calendar quarter; (12) keeps reaction sessions from 2020; (13) keeps events with ten or more pre-market bars on the morning of the base session, before any news; (14) keeps firms with one Item 2.02 filing in the quarter; (15) adds the minutes from the release to the end of its window as a control; (16) estimates the trend and the change on the WPC of each reaction quarter, without controls, with Newey--West standard errors (four lags); (17) subtracts beta times SPY's return from each window instead of SPY's return, with beta estimated from the 250 daily returns before the base session; (18) uses no benchmark; (19a) and (19b) split events by the sign of $X$; (20) dates the two event dummies by reaction session, 16 May to 15 November 2023 and 5 to 16 August 2024; (21) starts the after-hours window at 16:05, after the closing auction, which affects column (1) only; (22) drops $|X|$ from the controls. Standard errors otherwise as in Table~\ref{tab:tests}. Under each label, $N$ gives the numbers of post-close and pre-open events.}
\label{tab:robust2}
\end{table}

\subsection{H2: firm size and the after-hours share}

The level of the after-hours share differs by firm size: it is 0.48 for S\&P 500 firms, 0.41 for
S\&P 400 firms and 0.38 for S\&P 600 firms. The trends do not differ. The S\&P 600 trend is 0.0018
per year below that of the S\&P 500, with a standard error of 0.0085 (Table~\ref{tab:tests},
Panel~B), so the test does not support H2. Since the S\&P 600 enters only in 2020, we re-estimate
the tier model on reaction sessions from 2020; the difference is $-0.0014$ (0.0099). The minimum
detectable difference is about 0.024 per year. The size terciles are formed within each year, so
the bottom tercile contains mid caps before 2020 and small caps afterwards. Re-estimated from 2020,
its trend differs from that of the top tercile by 0.0047 (0.0096). Only the S\&P 400 and the middle
size tercile have higher trends, by about 0.013 per year, significant at the \SI{5}{\percent}
level. Among the 18 interaction terms in Panel~B, chance alone would produce about one significant
coefficient at the \SI{5}{\percent} level, and these two cover largely the same firms, so we do not
interpret them as a finding.

\subsection{H3: information in the opening price}

In every year and for both timings, the slope of $X$ on the return to 09:30 lies between 0.94 and
1.06, and the $R^2$ between 0.50 and 0.69 (Table~\ref{tab:open}). The $R^2$ has no trend ($+0.002$
per year, standard error 0.007), and neither has the winsorised $R^2$. With official prices the
$R^2$ is higher, at 0.52 to 0.72, since the official open is not stale, and it has no trend either
($+0.001$ and $+0.003$ per year, standard errors 0.007). For releases before the open, the slope
declines slightly from 1.04 in 2017 to 0.98 in 2026 ($-0.008$ per year, standard error 0.004); the
event-level test gives $-0.0063$ (0.0038), significant at the \SI{10}{\percent} level. For releases
after the close, the price at 20:00 already explains 0.40 to 0.59 of the variance of $X$, and the
next morning adds 0.08 to 0.15.

The test does not support H3 as stated, but it has little power. The slope was already 1 in 2017,
so it could not move toward 1. With ten yearly values, the $R^2$ trend has a standard error of
0.007, so the minimum detectable change is about 0.02 per year, or 0.2 over the decade. The share
priced by the open measures the same idea more precisely, and it rose slightly from 2016--2019 to
2024--2026. A larger share is consistent with a flat $R^2$ if the extended hours now account for
more of the average move but also add more noise, since thin trading prices the news with error;
the data do not allow us to separate these two effects.

\begin{table}[htbp]
\centering\scriptsize\setlength{\tabcolsep}{2pt}
\begin{tabular}{lrrrrrrrrrrr}
\toprule
& \multicolumn{6}{c}{After close} & \multicolumn{5}{c}{Before open} \\
\cmidrule(lr){2-7}\cmidrule(lr){8-12}
& & $R^2$, & Slope, & $R^2$, & $R^2$, 09:30, & $R^2$, & & Slope, & $R^2$, & $R^2$, 09:30, & $R^2$, \\
Year & Events & 20:00 & 09:30 & 09:30 & winsorised & open & Events & 09:30 & 09:30 & winsorised & open \\
\midrule
2017 & 736 & 0.59 & 1.00 & 0.69 & 0.68 & 0.72 & 909 & 1.04 & 0.67 & 0.64 & 0.67 \\
2018 & 920 & 0.43 & 0.94 & 0.56 & 0.55 & 0.62 & 997 & 1.04 & 0.59 & 0.57 & 0.63 \\
2019 & 931 & 0.54 & 0.94 & 0.64 & 0.63 & 0.70 & 1{,}091 & 1.00 & 0.65 & 0.63 & 0.69 \\
2020 & 1{,}468 & 0.42 & 0.97 & 0.53 & 0.51 & 0.59 & 1{,}339 & 1.06 & 0.50 & 0.51 & 0.52 \\
2021 & 1{,}637 & 0.40 & 0.97 & 0.50 & 0.51 & 0.54 & 1{,}508 & 1.01 & 0.53 & 0.54 & 0.54 \\
2022 & 1{,}642 & 0.47 & 0.96 & 0.55 & 0.52 & 0.63 & 1{,}431 & 0.94 & 0.55 & 0.55 & 0.59 \\
2023 & 1{,}874 & 0.47 & 1.04 & 0.62 & 0.58 & 0.68 & 1{,}653 & 0.98 & 0.64 & 0.62 & 0.68 \\
2024 & 2{,}032 & 0.51 & 0.98 & 0.61 & 0.60 & 0.69 & 1{,}881 & 1.02 & 0.63 & 0.62 & 0.67 \\
2025 & 2{,}170 & 0.52 & 1.00 & 0.65 & 0.63 & 0.68 & 2{,}186 & 0.96 & 0.59 & 0.59 & 0.62 \\
2026 & 1{,}802 & 0.50 & 0.94 & 0.64 & 0.62 & 0.66 & 1{,}673 & 0.98 & 0.67 & 0.62 & 0.68 \\
\midrule
Trend per year & & & +0.001 & +0.002 & +0.001 & +0.001 &  & $-$0.008 & +0.002 & +0.002 & +0.003 \\
 & & & (0.004) & (0.007) & (0.007) & (0.007) &  & (0.004) & (0.007) & (0.005) & (0.007) \\
Event-level slope trend & & & 0.0018 & & & & & $-$0.0063$^{*}$ & & & \\
 & & & (0.0038) & & & & & (0.0038) & & & \\
\bottomrule
\end{tabular}
\caption{H3: the share of the event return priced at each time. For each year and release timing, OLS of $X$ on the return from the last close before the release to 20:00 or to 09:30. A slope near 1 and a high $R^2$ indicate that the price at that time already reflects the news. For releases before the open, the return to 20:00 ends before the release and is not shown. The winsorised $R^2$ limits both returns to the year's 1st and 99th percentiles. The last column of each block uses official prices only: the $R^2$ of a regression of the return between two official closes on the return from the base close to the official open. The trend is an OLS slope on the ten yearly values, with its standard error in parentheses. The event-level slope trend is the coefficient on the interaction of the 09:30 return and the trend in a pooled regression of $X$ on the 09:30 return, the trend and their interaction, with standard errors clustered by firm and release date. December 2016 has too few events for a separate estimate.}
\label{tab:open}
\end{table}

\subsection{October and November 2026}

For the 454 releases after the close in October and November 2026, the after-hours session
accounts for 0.34 of the move (\SI{95}{\percent} interval 0.26 to 0.44). This value is low relative
to full years, but releases in October and November have a lower after-hours share in most years:
0.36 in 2021 and 0.35 in 2023 (Table~\ref{tab:season}). The 2026 value is 0.01 below the lowest
value of the earlier autumns, and its interval overlaps those of all nine earlier autumns. The data
contain overnight trades for \SI{23}{\percent} of these releases in 2025 and \SI{2}{\percent} in
2026, which reflects the change in LSE Data's coverage.

\begin{table}[htbp]
\centering\footnotesize\setlength{\tabcolsep}{3.5pt}
\begin{tabular}{lrrrr}
\toprule
& & \multicolumn{2}{c}{After-hours WPC} & With an \\
\cmidrule(lr){3-4}
Year & Events & Estimate & 95\% interval & overnight bar \\
\midrule
2017 & 181 & 0.51 & [0.39, 0.62] & 0\% \\
2018 & 234 & 0.45 & [0.36, 0.56] & 0\% \\
2019 & 200 & 0.44 & [0.34, 0.52] & 0\% \\
2020 & 350 & 0.40 & [0.33, 0.48] & 0\% \\
2021 & 395 & 0.36 & [0.31, 0.41] & 0\% \\
2022 & 361 & 0.40 & [0.34, 0.46] & 2\% \\
2023 & 454 & 0.35 & [0.29, 0.42] & 6\% \\
2024 & 484 & 0.45 & [0.41, 0.50] & 16\% \\
2025 & 534 & 0.44 & [0.40, 0.47] & 23\% \\
2026 & 454 & 0.34 & [0.26, 0.44] & 2\% \\
\bottomrule
\end{tabular}
\caption{After-hours share (WPC) of releases after the close whose reaction session falls in October or November, under the rules of the main sample, and the share of these releases with at least one bar from 00:00 to 04:00 in the data. 2026 runs to 24 November. Intervals are 95\% bootstrap percentile intervals from 2,000 resamples of release dates.}
\label{tab:season}
\end{table}

\FloatBarrier
\section{Discussion}\label{sec:discussion}

The main limitation of this study is that a single data source, which records only part of the
market, measures both the market and its own coverage. The next-morning result shows the cost of
this limitation: an increase that held in a constant panel, with fresh prices and among stocks with
dense trading before the news, fell by half when measured against official opening prices.

\paragraph{Coverage.} LSE Data records \SI{44}{\percent} of the regular session's volume in every
year, but its pre-market coverage increases: in the main sample, the share of stocks with ten or
more pre-market bars on a morning before the news doubled. Official opening prices fix the price at
09:30 independently of the bars, and LSE Data's own first trade from 09:30 agrees with them. The
price at 20:00 has no such check. If the last trade before 20:00 is stale, part of the evening's
move is assigned to the next morning, so the split between the two windows depends on the data
alone. The volume ratio may also conceal a change. The bars from 16:00 to 16:05, which contain the
closing auction, account for \SI{6}{\percent} of the data's regular-session volume in 2017 and
\SI{15}{\percent} in 2026. Closing auctions grew across the market
\citep{bogousslavsky-muravyev-2023}. If LSE Data's share of these auctions grew as well, its
coverage of continuous trading fell while the total remained at \SI{44}{\percent}. A comparison with the
consolidated tape for a sample of event days would resolve both questions; we have not performed
it.

\paragraph{Trades and quotes.} The windows end at the last trade, and the data contain no quote
history. For releases after the close, the last trade before 20:00 falls in the last five minutes
for only 30 to \SI{39}{\percent} of events, with no trend; for the last trade before 09:30, the
figure is 53 to \SI{67}{\percent}, higher in 2025 and 2026 (Table~\ref{tab:activity}). Restricting
the sample to events with fresh prices at both ends (check 8) did not reveal the stale-price problem
at 09:30, because it keeps a selected third of the sample.

\paragraph{Prices and benchmark.} The base price is the last trade before 16:00, so the move to the
closing auction, 0.03 of the event return, falls in the after-hours window; check 21 removes it. The
Yahoo Finance check bounds the difference between the two-day return and the official return at
\SI{0.5}{\percent}. Returns are net of SPY with a beta of one. With a beta estimated for each firm,
or with no benchmark, the after-hours change is slightly larger and the other results are unchanged
(checks 17 and 18). We have not used a small-cap benchmark for small firms.

\paragraph{No measure of the news.} The tests control for the size of the move but not for the
earnings surprise or for other content of the filing, and they include no control days around each
event. The share priced by the open on days without news provides a first comparison. A complete
comparison would measure every window on such days, which requires more bars.

\paragraph{Survivorship and selection.} Firms delisted before 2026 are missing; they account for a
fifth of the 2017 releases. The constant panel of firms with a release in every year (check 6b),
which contains only survivors throughout, gives the same results. The coverage rule keeps events
with a trade in the data after the EDGAR time. Without it, the after-hours change is larger and the
pre-market trend measured to the last trade doubles (check 1). The share priced by the official open
does not need the rule, and removing it changes this share little.

\paragraph{Release times.} For many firms, the EDGAR acceptance comes after the newswire release.
Half of the after-hours response after 16:05 precedes the EDGAR time, and in 29 to \SI{37}{\percent}
of main-sample events the price completes half of its move before that time. The windows do not
depend on the exact release time, but the coverage rule and check 15 do. Excluding releases after
the close whose first half-way move comes before 16:05 changes no result (check 2). No release
before the open changes timing class in this way, since a move up to an hour earlier still falls
before the open.

\paragraph{Small caps.} The S\&P 600 enters in 2020, so H2 compares tiers over seven years, and the
pooled yearly shares combine two universes; where this matters, we report S\&P 500 and 400 firms
separately.

\paragraph{Order of work and multiple tests.} We computed the year-by-year shares before the exact
tests were specified. Checks 8 to 16 were added after the main results, and checks 17 to 22 and the
official-open tests after a second review. We report 22 checks on three windows, and some
coefficients will be significant by chance, so we base the conclusions on patterns across checks
rather than on individual significance levels. The first draft interpreted the next-morning increase
as a change in the market. The comparison with official prices, which was not among the planned
checks, showed that half of the increase came from stale prices.

\paragraph{Replication.} The bars are licensed for personal research and cannot be shared. LSE Data
serves only a rolling ten-year window, so the earliest months will soon no longer be available. The
tests on the share priced by the open use only public daily prices and EDGAR times, so anyone can
replicate them. A replication of the other results on the consolidated tape would avoid both
problems.

\section{Conclusion}\label{sec:conclusion}

Trading outside regular hours grew from 2016 to 2026, but the after-hours share of the earnings
response did not grow with it. The share fell from 0.52 in 2017 to 0.38 in 2021 and is 0.43 in
2026. Its change from 2016--2019 to 2024--2026 is between 0.01 and 0.04, depending on the benchmark
and on whether the closing auction is included, and the data rule out an increase of more than 0.10
to 0.13 over the decade. In the data, after-hours trades became more numerous and much smaller, but
their share of total volume did not increase.

Half of the apparent doubling of the next-morning share is due to stale prices. With official
opening prices, the share of the response priced by the open is 0.05 to 0.07 higher in 2024--2026
than in 2016--2019, and the same share rose by 0.04 on days without earnings news. What
distinguishes earnings releases is a decline in 2020--2023, when the regular session accounted for
more of the response to releases after the close. Small firms follow the same path as large firms,
and the opening price predicts the next close as well as before.

Three questions remain open: why the regular session accounted for more of the response in
2020--2023, whether the consolidated tape confirms the split between the evening and the next
morning, and how the end of overnight trades in the data in October 2026 affects the after-hours
share.

\section*{Declarations}

\paragraph{Data availability.} The 1-minute bars come from LSE Data \citep{lse-data} under a
licence for personal research use and are not redistributed. EDGAR filings and XBRL share counts
are publicly available from the SEC. Index membership comes from the public page histories of
Wikipedia's lists of S\&P 500, S\&P 400 and S\&P 600 companies, daily opening and closing prices
and volumes from Yahoo Finance, and VIX data from Cboe. The aggregate tables and the yearly window
shares behind Figure~\ref{fig:wpc} are stored with the code.

\paragraph{Code availability.} The analysis consists of scripts in a private repository, one for
each step from the downloaded data to the tables and figure of this paper. The tables and the
figure are generated by a script and are not edited by hand.

\paragraph{Use of automated assistance.} The author used large language models as coding and
writing assistants for the analysis code, the text, the figures and reviews of drafts.

\paragraph{Funding and competing interests.} The author received no funding for this work and
declares no competing interests.
\clearpage
\bibliographystyle{plainnat}
\bibliography{refs}

\clearpage
\appendix
\section{Validation}\label{app:validation}

We checked the release times and LSE Data's prices before examining any result.

\paragraph{Release times are in UTC.} In every year from 2017, 89 to \SI{95}{\percent} of EDGAR
acceptance times fall between 06:00 and 09:00 ET or between 16:00 and 20:00 ET.

\paragraph{Release times match prices.} In the main sample, the median time from acceptance to the
first half-way move is within 10 minutes of zero in every year and for both timings, except for
releases before the open in 2017 (16 minutes). In 29 to \SI{37}{\percent} of events, the price
completes half of its move before the EDGAR time, since newswire releases often come first. Timing
these releases by the price move assigns 684 events to another timing class.

\paragraph{Bars are available.} Every event has after-hours and regular-session bars, and 99.7 to
\SI{99.9}{\percent} of events pass the price checks. Of all events with bars, \SI{66}{\percent}
have a pre-market trade on the reaction morning in 2017, \SI{61}{\percent} in 2022,
\SI{72}{\percent} in 2024 and \SI{87}{\percent} in 2026; in the main sample, which requires a trade
after the release, the figure is 83 to \SI{93}{\percent}. The main sample has no overnight bars
before 2022; \SI{1}{\percent} of its events have one in 2022, \SI{5}{\percent} in 2023,
\SI{15}{\percent} in 2024, \SI{19}{\percent} in 2025 and \SI{28}{\percent} in 2026.

\paragraph{Prices match Yahoo Finance.} LSE Data's two-day event return agrees with Yahoo Finance's
close-to-close return within \SI{0.5}{\percent} for \SI{98.2}{\percent} of the events for which
Yahoo Finance has prices. We exclude the 696 events that do not agree; 462 of them fall in
2020--2022, and the median discrepancy is 66 basis points. The 239 events without a Yahoo Finance
price remain in the sample. A filter for isolated 15:59 prints that differ by more than
\SI{2}{\percent} from both neighbouring prints removes 3 further events.

\paragraph{Official opens match the bars.} The return from the base price to Yahoo Finance's
official open is within \SI{0.5}{\percent} of the return to LSE Data's first trade from 09:30 for
\SI{88}{\percent} of main-sample events, and the median difference is 8 basis points; the first
trade need not be part of the opening auction. The two prices give the same shares in every year
(Table~\ref{tab:official}, last rows).

\paragraph{Volumes are a stable fraction.} At the median, LSE Data's volume from 09:30 to 16:05 on
the base session is \SI{44}{\percent} of Yahoo Finance's daily volume in every year and every index
tier, and it lies between 42 and \SI{45}{\percent} for half of all events.

\paragraph{Tickers match EDGAR.} Every LSE Data symbol is in EDGAR's ticker list for its firm,
except for three share-class spellings (BF.B, BRK.B, MOG.A).

\paragraph{Large moves are genuine where checked.} In the main sample, 27 events move by more than
\SI{50}{\percent}. Those we checked are known earnings-day crashes, such as DexCom in July 2024 and
Synopsys in September 2025.

\paragraph{Market capitalisations are plausible.} At the prior quarter end, Apple's market
capitalisation is \$618 billion and NVIDIA's \$58 billion in early 2017; \SI{98.5}{\percent} of
events have a market capitalisation.

\end{document}